\documentclass[showpacs,aps,preprint,amsfonts]{revtex4}

\usepackage{epsfig}
\usepackage{amsmath}
\usepackage{bm}

\begin{document}

\title{The evolution with temperature of magnetic polaron state
 in an antiferromagnetic chain with impurities}

\author{A.~O.~Sboychakov, A.~L.~Rakhmanov, and K.~I.~Kugel}
\affiliation{Institute for Theoretical and Applied
Electrodynamics, Russian Academy of Sciences, Izhorskaya Str.
13/19, Moscow, 125412 Russia}

\author{I.~Gonz\'alez, J.~Castro, and D.~Baldomir}
\affiliation{Departamento de F\'{\i}sica Aplicada, Universidade de
Santiago de Compostela, E-15706 Santiago de Compostela, Spain}


\begin{abstract}

The thermal behavior of a one-dimensional antiferromagnetic chain
doped by donor impurities was analyzed. The ground state of such a
chain corresponds to the formation of a set of ferromagnetically
correlated regions localized near impurities (bound magnetic
polarons). At finite temperatures, the magnetic structure of the
chain was calculated simultaneously with the wave function of a
conduction electron bound by an impurity. The calculations were
performed using an approximate variational method and a Monte
Carlo simulation. Both these methods give similar results. The
analysis of the temperature dependence of correlation functions
for neighboring local spins demonstrated that the ferromagnetic
correlations inside a magnetic polaron remain significant even
above the N\'eel temperature $T_N$ implying rather high stability
of the magnetic polaron state. In the case when the
electron-impurity coupling energy $V$ is not too high (for $V$
lower that the electron hopping integral $t$), the magnetic
polaron could be depinned from impurity retaining its magnetic
structure. Such a depinning occurs at temperatures of the order of
$T_N$. At even higher temperatures ($T \sim t$) magnetic polarons
disappear and the chain becomes completely disordered.

\end{abstract}


\maketitle

\section{Introduction}

The tendency to formation of inhomogeneous charge and spin states
and to phase separation is of fundamental importance for the
physics of manganites and other systems with strongly correlated
electrons. It is now a common belief that the nature of the
colossal magnetoresistance effect is closely related to phase
separation phenomenon~\cite{dag03b}. The most widely discussed
type of phase separation is the formation of small magnetic
droplets (magnetic polarons or ferrons) as a result of the
self-trapping of charge carriers in an insulating
antiferromagnetic matrix. This type of phase separation, first
discussed by Nagaev in his analysis of electron states in magnetic
semiconductors~\cite{nag67}, is now actively used for
interpretation of experimental data on manganites at different
doping levels~\cite{dag03b,sal01}.

In this connection, a problem arises, whether these self-trapped
electron states can survive above the characteristic temperature
of the magnetic ordering. The possible existence of the
ferromagnetic correlated regions at the paramagnetic background,
``temperature ferrons'', was first discussed in detail by
Krivoglaz~\cite{kri74} (see also \cite{kag01}). Indeed, there are
a lot of experimental indications (coming mostly from the
susceptibility and ESR data) that manganites in the paramagnetic
state are rather inhomogeneous. Moreover, the analysis of the
high-temperature transport properties of manganites based on the
assumption of the existence of ferromagnetically correlated
regions in the paramagnetic phase gives a viable picture of the
inhomogeneous state~\cite{kug04}.

In our paper, we study the evolution with temperature of a magnetic polaron
state over a wide temperature range using a simple model such as an
antiferromagnetic chain doped by donor impurities. In spite of its
simplicity, this model reveals the possible existence of different kinds
of ferrons~\cite{art3,art5} and seems to capture the essential features of
the electronic phase separation in manganites at low doping levels. At low
doping, the impurity potential is a relevant parameter and we start
from the state corresponding to ferrons bound by impurities. The density of
impurities is assumed to be small, so that ferrons do not overlap, and the
system as a whole remains insulating. We demonstrate that ferrons are
rather stable objects and ferromagnetic correlations persist at
temperatures much higher than the N\'eel temperature of the host chain.

The structure of the paper is as follows. In Section~\ref{sec2},
we formulate our model for the antiferromagnetic chain doped by
donor impurities. In Section~\ref{sec3}, two different methods to
calculate the partition function for such a chain, using an
approximate variational method and a Monte Carlo simulation, are
presented. Finally, in Section~\ref{sec4}, we discuss the
characteristic features of the model in different temperature
ranges.

\section{\label{sec2}Formulation of the model}

We consider a one-dimensional chain of antiferromagnetically coupled
local spins. Non-magnetic donor impurities are distributed homogeneously
with a period $L$ (in lattice constant units) along the chain. The system is
described by the double-exchange Hamiltonian,

\begin{equation}\label{eq:1a}
H =
H_{\text{el}}+J'\sum\limits_{g}\vec{S_{g}}\cdotp\vec{S}_{g+1}\,,
\end{equation}
\begin{eqnarray}\label{eq:1b}
H_{\text{el}} &=& -t\sum\limits_{g,s}
\left(a^{\dag}_{g,s}a_{g+1,s}+a^{\dag}_{g+1,s}a_{g,s}\right)-
\nonumber\\ & & -\frac{A}{2} \sum\limits_{g,s,s'}
a^{\dag}_{g,s}\left(\vec{\sigma}\cdotp
\vec{S}_{g}\right)_{s,s'}
a_{g,s'}+H_{\text{imp}}\,, \\
H_{\text{imp}}&=&-\sum\limits_{g,s} V_{g}a^{\dag}_{g,s}a_{g,s}\,.
\end{eqnarray}
In Eqs.~(\ref{eq:1a}) and (\ref{eq:1b}), $\vec{S_{g}}$ is the spin
of the magnetic ion located at site $g$, treated as a classical
vector, symbols $a^{+}_{g,s}$, $a_{g,s}$ denote the creation and
annihilation operators for the conduction electron with spin
projection $s$ at site $g$, and $\vec{\sigma}$ are Pauli matrices.
The second term in Eq.~(\ref{eq:1a}) is the antiferromagnetic
exchange between local spins. The two terms in $H_{\text{el}}$
describe the kinetic energy of conduction electrons bounded by
impurities, which are located between sites with $g=iL$ and
$g=iL+1$ ($i$ is an integer), and the Hund's-rule coupling between
the conduction electrons and the local spins. $H_{\text{imp}}$ is
the electrostatic interaction between an impurity and conduction
electron. Parameters $V_{g}$, $A$, $t$, and $J'$ of
Hamiltonian~(\ref{eq:1a}) are considered to be positive. The
hierarchy of this parameters is $A>t, V_{g}\gg J'$, as usual in
the double exchange approximation.

As we pointed out in Introduction, the electron-impurity
interaction term is needed to describe the low-doping limit of the
model, when the material is insulating. In our previous
discussions on this model~\cite{art3,art5}, this potential was
chosen as a deep square well. Although this is a good
approximation for $T=0$, here we prefer to use a more realistic
form for this interaction, namely the Coulomb potential. This
allows us to find self-consistently the wave function, and hence
the size of ferron, which becomes temperature-dependent. In fact,
the calculations show that results are not strongly affected by
the specific form of the impurity potential.

We consider the low-doping limit of the model, which corresponds
to large $L$. This allows us to neglect the interaction between
electrons corresponding to different impurities, and also to
restrict our calculation to a finite portion of the chain. In the
part of the chain under study, of length $L$, there just one
impurity located in its middle between two magnetic sites.

Hamiltonian~(\ref{eq:1a},\ref{eq:1b}) is rotationally invariant,
so we can choose angles $\nu_{g}$ between two neighboring spins in
the chain as variables characterizing the magnetic structure of
the local spins. In the limit $A\rightarrow \infty$, only one of
the spin components contributes to the low-lying states. Starting
from operators $a^{\dag}_{g,s}$, we make a transformation to
operators $c^{\dag}_{g}$, which describe the creation of a
conduction electron with its spin directed along each local spin
$S_{g}$ (spinless fermions). This modifies the hopping term that
now depends on the magnetic structure of local spins. Using angles
$\nu_{g}$, we can write Hamiltonian~(\ref{eq:1a}) as:

\begin{eqnarray}\label{eq:2}
H&=&H_{\text{el}}+ J\sum_{g=-L/2}^{L/2-2}\cos\nu_{g}\nonumber\,,\\
H_{\text{el}}&=&-t\sum_{g=-L/2}^{L/2-2}
\left(c^{\dag}_{g+1}c_{g}+c^{\dag}_{g}c_{g+1}\right)
\cos\frac{\nu_{g}}{2}
-V\sum_{g=-L/2}^{L/2-1}\frac{\displaystyle{c^{\dag}_{g}c_{g}}}
{\left|g+1/2\right|}\,,
\end{eqnarray}
where the Coulomb form of the impurity potential is introduced
explicitly, and $J=J'S^{2}$.

\section{\label{sec3}Calculation of the partition function}

Using Hamiltonian~(\ref{eq:2}), we can write the partition
function in the following form:

\begin{equation}\label{Z}
Z=\sum_{n=0}^{L-1}\left\{\prod_{g=-L/2}^{L/2-2}\left(\int\limits_{0}^{\pi}
d\nu_{g}\sin\nu_{g}e^{\displaystyle{-\beta J\cos\nu_{g}}}\right)
\left\langle\psi_{n} \left|\right.\right.e^{\displaystyle{-\beta
H_{\text{el}}}}\left|\psi_{n}\left\rangle\right.\right.\right\}\ ,
\end{equation}
where $\beta=1/T$, and $\left|\psi_{n}\left\rangle\right.\right.$
are the basis electron wave functions, which are assumed to be
orthogonal:
\begin{equation}\label{ort}
\left|\psi_{n}\left\rangle\right.\right.=\sum_{g=-L/2}^{L/2-1}
\psi_{g}^{(n)}c^{\dag}_{g}\left|0\left\rangle\right.\right.\ , \text{
} \sum_{g=-L/2}^{L/2-1}\psi_{g}^{(m)}\psi_{g}^{(n)}=\delta_{n,m}\ .
\end{equation}
In this expression, we suppose that $\psi_{g}^{(n)}$ are real
numbers, because the matrix elements of operator
$H_{\text{el}}$ in the
$c^{\dag}_{g}\left|0\left\rangle\right.\right.$ basis are real. Note that
we can use any appropriate basis to calculate the partition
function. If $\left|\psi_{n}\left\rangle\right.\right.$ are the
eigenfunctions of the operator $H_{\text{el}}$, the last factor in
Eq.~(\ref{Z}) can be written in the form:
\begin{equation}
\left\langle\psi_{n} \left|\right.\right.e^{\displaystyle{-\beta
H_{\text{el}}}}\left|\psi_{n}\left\rangle\right.\right.=
e^{\displaystyle{-\beta\left\langle\psi_{n} \left|\right.\right.
H_{\text{el}}\left|\psi_{n}\left\rangle\right.\right.}}=
e^{\displaystyle{-\beta E_{n}}}\ .
\end{equation}
In this case, wave functions
$\left|\psi_{n}\left\rangle\right.\right.$ depend, of course, on
the canting angles.

\subsection{Approximate partition function}

Partition function~(\ref{Z}) can not be calculated
analytically. In this subsection, we propose an approximate
approach to the evaluation of the partition function and physical
observables. The main point of this approach is the substitution
in Eq.~(\ref{Z}):

\begin{equation}\label{sub}
\left\langle\psi_{n} \left|\right.\right.e^{\displaystyle{-\beta
H_{\text{el}}}}\left|\psi_{n}\left\rangle\right.\right.\rightarrow
e^{\displaystyle{-\beta\left\langle\overline{\psi}_{n}
\left|\right.\right.
H_{\text{el}}\left|\overline{\psi}_{n}\left\rangle\right.\right.}}
\ ,
\end{equation}
where effective wave functions
$\left|\overline{\psi}_{n}\left\rangle\right.\right.$ do not
depend on canting angles $\nu_{g}$. The procedure of finding these
wave functions is described below. The accuracy of such a
procedure is tested by the Monte Carlo simulations. The physical
meaning of this substitution is that the main contribution to $Z$
comes from the term corresponding to the ground state of the
conduction electron. Since each electron in the ground state is
trapped by its impurity, we can conclude that the its wave
function only slightly depends on the canting angles.

After substitution of Eq.~(\ref{sub}) into Eq.~(\ref{Z}), the approximate
partition function can be written in the form:

\begin{eqnarray}
Z^{\text{appr}}&=&
\sum_{n=0}^{L-1}\exp\left[\displaystyle{\beta V\sum_{g=-L/2}^{L/2-1}
\frac{\left(\overline{\psi}^{(n)}_{g}\right)^2}{|g+1/2|}}\right]\times
\nonumber\\
& &\prod_{g=-L/2}^{L/2-2}\left\{\int\limits_{0}^{\pi}
d\nu_{g}\sin\nu_{g}
\exp\left[\displaystyle{\beta\left(2t\overline{\psi}^{(n)}_{g+1}
\overline{\psi}^{(n)}_{g}\cos\frac{\nu_{g}}{2}-J\cos\nu_{g}\right)}\right]\right\}
\ ,
\end{eqnarray}
where we use Eq.~(\ref{eq:2}) for operator $H_{\text{el}}$.

The wave functions $\overline{\psi}^{(n)}_{g}$ are found by
minimization of the approximate free energy $F^{\text{appr}}=-T\ln
Z^{\text{appr}}$ with respect to $\overline{\psi}^{(n)}_{g}$.
However, we should take into account that wave functions
$\overline{\psi}^{(n)}_{g}$ are not independent because of the
orthogonality conditions~(\ref{ort}). Calculating the derivatives
$\partial F^{\text{appr}}/\partial\overline{\psi}^{(n)}_{g}$, we
obtain the following system of nonlinear equations:

\begin{equation}\label{psi}
\left(\langle\hat{H_{\text{el}}}\rangle_{(n,n)}\overline{\psi}^{(n)}\right)_{g}
-\sum_{m=0}^{n}\Lambda_{n,m}\overline{\psi}^{(m)}_{g}=0\ ,\text{ }
n=0,1, \dots L-1\ ,
\end{equation}
where $\hat{H_{\text{el}}}$ is the matrix of the operator
$H_{\text{el}}$ in the
$c^{\dag}_{g}\left|0\left\rangle\right.\right.$ basis ,
$\Lambda_{n,m}$ ($m\leq n$) are the Lagrange multipliers, and
symbol $\langle\dots\rangle_{(n,m)}$ denotes the following
averaging procedure:

\begin{equation}\label{mean}
\left\langle f(\nu_{g})\right\rangle_{(n,m)}=
\frac{\displaystyle{\int\limits_{0}^{\pi}
d\nu_{g}\sin\nu_{g}f(\nu_{g})e^{\displaystyle{\beta\left[t\left(\overline{\psi}^{(n)}_{g+1}
\overline{\psi}^{(m)}_{g}+\overline{\psi}^{(m)}_{g+1}
\overline{\psi}^{(n)}_{g}\right)\cos\frac{\nu_{g}}{2}-J\cos\nu_{g}\right]}}}}
{\displaystyle{\int\limits_{0}^{\pi}
d\nu_{g}\sin\nu_{g}e^{\displaystyle{\beta\left[t\left(\overline{\psi}^{(n)}_{g+1}
\overline{\psi}^{(m)}_{g}+\overline{\psi}^{(m)}_{g+1}
\overline{\psi}^{(n)}_{g}\right)\cos\frac{\nu_{g}}{2}-J\cos\nu_{g}
\right]}}}}\ .
\end{equation}
The system of equations~(\ref{psi}) with additional conditions~(\ref{ort})
is solved numerically. Note that the approximate free
energy satisfies the inequality $F^{\text{appr}}>F$, because for
any canting angle the following condition is met
\begin{equation}
\left\langle\overline{\psi}_{n} \left|\right.\right.
H_{\text{el}}\left|\overline{\psi}_{n}\left\rangle\right.\right.\geq
E_n\ .
\end{equation}

Let us make some remarks concerning Eqs.~(\ref{psi}) and
(\ref{mean}). It can be easily seen from these equations that
parameter $\Lambda_{n,n}$ is the mean energy of the electron in
the $n$th state, that is
$\Lambda_{n,n}=\left\langle\overline{\psi}_{n}
\left|\right.\right. \langle\hat{H_{\text{el}}}\rangle_{(n,n)}
\left|\overline{\psi}_{n}\left\rangle\right.\right.$. We can say
that $\left|\overline{\psi}_{n}\left\rangle\right.\right.$ are the
states of an electron moving in the averaged local spin background
corresponding to these states. But, strictly speaking, only the
state with $n=0$ (``ground'' state) is the eigenstate of the
averaged operator $\langle H_{\text{el}}\rangle_{(0,0)}$.
\footnote{It can be shown that the state with $n=1$ is also an
eigenstate of $\langle H_{\text{el}}\rangle_{(1,1)}$. It follows
from the mirror symmetry of the Hamiltonian and quantum-mechanical
sign theorem.} It is clear that in the case when the difference
between energies $\Lambda_{n,n}$ is large enough in comparison
with temperature, that is
$\beta(\Lambda_{n+1,n+1}-\Lambda_{n,n})\gg1$, we can neglect
contributions to the free energy from all states with $n>0$.

The approximate approach described here allows us to calculate
also the physical observables. Let us consider first the local
spin sector of the problem. The mean value of the function
$F(\{\theta_{g}, \phi_{g}\})$ can be easily calculated only in the
special case when: 1) $F$ depends only on $\nu_{g}$, and 2)
it is possible to separate the variables in the $(L-1)$-dimensional
integrals over the canting angles. The simplest case is:
\begin{equation}
F(\{\theta_{g}, \phi_{g}\})=\sum_{g=-L/2}^{L/2-2}f(\nu_{g})\ .
\end{equation}
It is useful to introduce the notation:
\begin{eqnarray}
Z_{n,m}&=&
\exp\displaystyle{\left(\beta
V\sum_{g=-L/2}^{L/2-1}
\frac{\overline{\psi}^{(n)}_{g}\overline{\psi}^{(m)}_{g}}{|g+1/2|}\right)}\times\nonumber\\
& &\prod_{g=-L/2}^{L/2-2}\left\{\int\limits_{0}^{\pi}d\nu_{g}\sin\nu_{g}\exp
\displaystyle{\left[\beta t
\left(\overline{\psi}^{(n)}_{g+1}\overline{\psi}^{(m)}_{g}+
\overline{\psi}^{(m)}_{g+1}\overline{\psi}^{(n)}_{g}\right)\cos\frac{\nu_{g}}{2}
-\beta J\cos\nu_{g}\right]}\right\}.
\end{eqnarray}
The mean value of $f(\nu_{g})$ then reads:

\begin{equation}\label{meanF}
\left\langle f(\nu_{g})\right\rangle=
\frac{1}{Z^{\text{appr}}}
\sum_{n=0}^{L-1}Z_{n,n}\left\langle f(\nu_{g})\right\rangle_{(n,n)}\ .
\end{equation}

To calculate the mean value of the operator $O$ corresponding to
the physical quantity in the electron sector, we should made the
transformation:
\begin{equation}
\left\langle\psi_{n}\left|\right.\right.Oe^{\displaystyle{-\beta
H_{\text{el}}}}\left|\psi_{n}\left\rangle\right.\right.\rightarrow
\sum_{m=0}^{L-1}O_{n,m}
e^{\displaystyle{-\beta\left\langle\overline{\psi}_{m}
\left|\right.\right.
H_{\text{el}}\left|\overline{\psi}_{n}\left\rangle\right.\right.}}
\ ,
\end{equation}
where $O_{n,m}=\left\langle\overline{\psi}_{n}
\left|\right.\right.
O\left|\overline{\psi}_{m}\left\rangle\right.\right.$ are the
matrix elements of the operator $O$ in the basis
$\left|\overline{\psi}_{n}\left\rangle\right.\right.$. The mean
value of $O$ can be written in the following form:

\begin{equation}\label{meanO}
\left\langle O\right\rangle=\frac{1}{Z^{\text{appr}}}
\sum_{n,m=0}^{L-1}O_{n,m}Z_{m,n}\ .
\end{equation}

We can also find mean values of physical observables in more
complicated cases. For example, the approximate value of the mean
energy of the system is:

\begin{eqnarray}\label{meanE}
\left\langle E\right\rangle&=&-\frac{1}{Z^{\text{appr}}}\sum_{n,m=0}^{L-1}
\left\{Z_{m,n}\left[V\sum_{g=-L/2}^{L/2-2}\frac{\overline{\psi}^{(n)}_{g}
\overline{\psi}^{(m)}_{g}}{|g+1/2|}
+\right.\right.\nonumber\\
&+&\left.\left.
t\sum_{g=-L/2}^{L/2-2}\left(\overline{\psi}^{(n)}_{g+1}
\overline{\psi}^{(m)}_{g}+\overline{\psi}^{(m)}_{g+1}
\overline{\psi}^{(n)}_{g}\right)\left\langle
\cos\frac{\nu_g}{2}\right\rangle_{(n,m)}\right]\right\}\\
& & +\frac{J}{Z^{\text{appr}}}
\sum_{n=0}^{L-1}\sum_{g=-L/2}^{L/2-2}Z_{n,n}\left\langle
\cos\nu_g\right\rangle_{(n,n)}\nonumber\ .
\end{eqnarray}
At relatively low temperatures,
$\beta(\Lambda_{n+1,n+1}-\Lambda_{n,n})\gg1$,
formulas~(\ref{meanF}), (\ref{meanO}), and (\ref{meanE}) for mean
values have a simpler form. For example, the mean value of
$f(\nu_{g})$ is equal to $\left\langle
f(\nu_{g})\right\rangle_{(0,0)}$. To test the approximate approach
described above, we calculate also the partition function and
physical quantities characterizing the temperature behavior of the
bound magnetic polaron using Monte Carlo simulations.

\subsection{Monte Carlo simulations}

Partition function~(\ref{Z}) can be numerically calculated using
standard diagonalization subroutines and a Monte Carlo (MC) simulation.
For a given local spin configuration, the electronic Hamiltonian
is quadratic in the $c_{g}$ operators and can be exactly diagonalized. The
remaining integrals over the local spin configurations can be calculated
using a classical MC simulation. The MC simulation of the local spin
system is done using the Metropolis algorithm (see, for example,
Ref.~\cite{bin97}). The Metropolis algorithm generates a set of local spin
configurations according to some probability distribution for the local
spin angles. We use:

\begin{equation}
P\left(\left\{\nu_{g}\right\}\right)=
\prod\limits_{g=-L/2}^{L/2-2}
\left(e^{\displaystyle{-\beta J\cos \nu_{g}}}\right)
\sum\limits^{L-1}_{n=0}
e^{\displaystyle{-\beta E_{n}\left(\{\nu_{g}\}\right)}}\ .
\end{equation}
Additional weight factor coming from the volume elements of
integrals~(\ref{Z}) is included in the course of calculating the
physical observables. Note that for the diagonalization of the
electronic Hamiltonian, one needs to know the local spin
configuration, and for the MC simulation of the local spin system,
one needs to know the ground state of the electronic Hamiltonian.
We proceed as follows. Starting from a given spin configuration,
we diagonalize the electronic Hamiltonian, and calculate
$P\left(\{\nu_{\text{g}}\}\right)$. Then we choose a site $g$ at
random, and temporarily change $\nu_{g} \to \nu_{g}'$. The electronic
Hamiltonian is again diagonalized and the new
$P\left(\{\nu_{\text{g}}'\}\right)$ calculated. The change in
canting angles is finally accepted according to the Metropolis
algorithm.
We define a MC step as a number of these spin reorientations equal
to the number of canting angles, $L-1$. Typically, we use series
of several thousands of MC steps for each set of parameters of the
Hamiltonian and $T$. A half of these steps is used to thermalize
the system, and another half is used to calculate the physical
observables. Between each two MC steps, we discard 10 extra MC
steps to avoid correlations between consecutive measurements. For
the spin reorientations, the new value of the corresponding
canting angle is randomly chosen. Open boundary conditions are
used at the ends of the chain. Note that in contrast to the
approximate method described in the previous subsection, the MC
simulation can reveal some effects related to the dynamics of the
conduction electron, the time being defined by the MC steps.

\section{\label{sec4}Results}

We calculate partition function of the model (\ref{eq:2}) for the
chain with $16$ sites at different values of parameters $V$, $t$
and $J$. This length is representative of the low doping limit,
and it is found that further increase of it does not change the
obtained results. The impurity is placed in the middle of the
chain between sites with $g=-1$ and $g=0$. To compare results
given by variational method and MC simulation, we calculated the
mean energy of the system $E$ in the temperature range $0<T<2.5J$.
The calculations of function $E(T)$ for different parameters of
the model show that difference between both methods does not
exceed several percent. Some test MC calculations at higher
temperatures also demonstrate a good agrement with approximate
procedure. So, we believe that our variational approach is a good
approximation even at high temperatures, and calculate physical
quantities at $T>2.5J$ using this method only.

As was already mentioned in Section~\ref{sec2}, our methods allow
us to calculate the wave function for the conduction electrons.
From this, we can estimate the size of the bound ferron. We have
used different set of parameters of the Hamiltonian, and found
that the size of magnetic polaron depends only slightly on the
interaction between impurity and conduction electron, $V$. For
example, for $V=100J$ magnetic polaron contains $4$ magnetic
sites, whereas for $V=10J$ it extends over $6$ sites.

\begin{figure}
\begin{center}
\epsfig{file=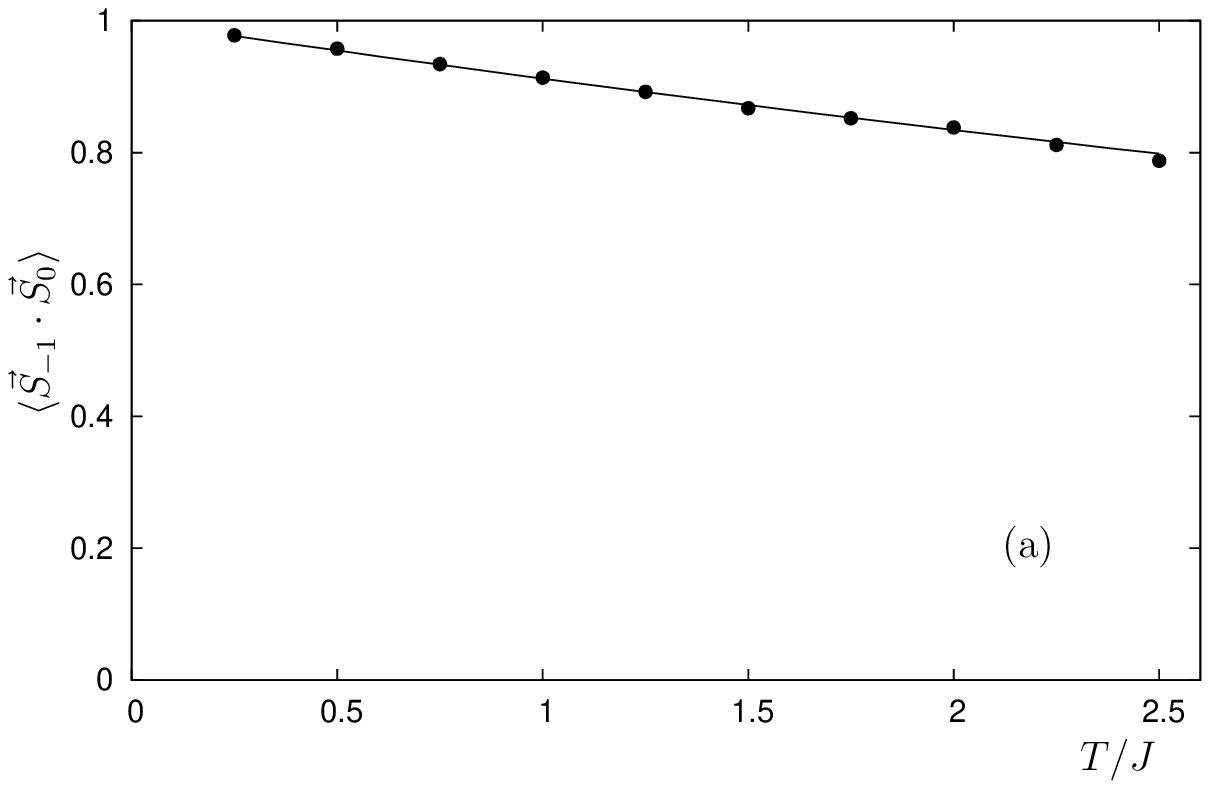,width=0.45\textwidth}
\epsfig{file=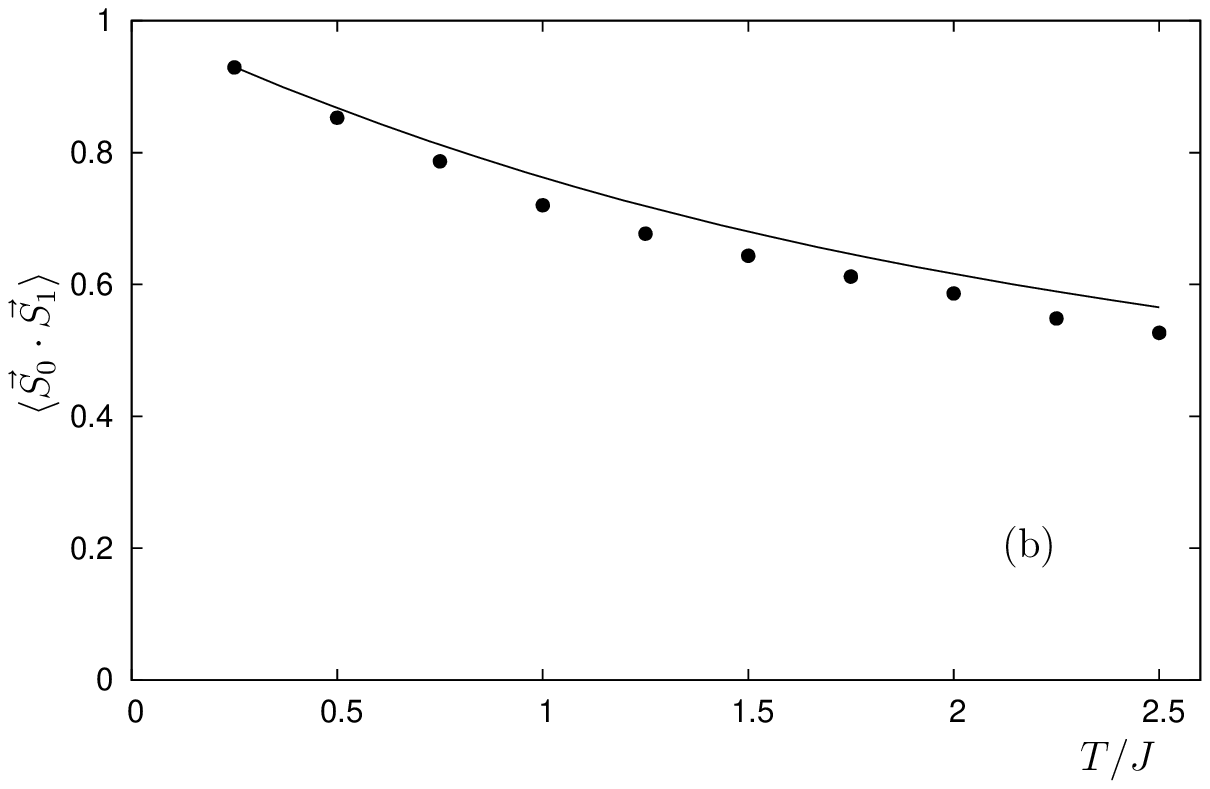,width=0.45\textwidth}\\
\epsfig{file=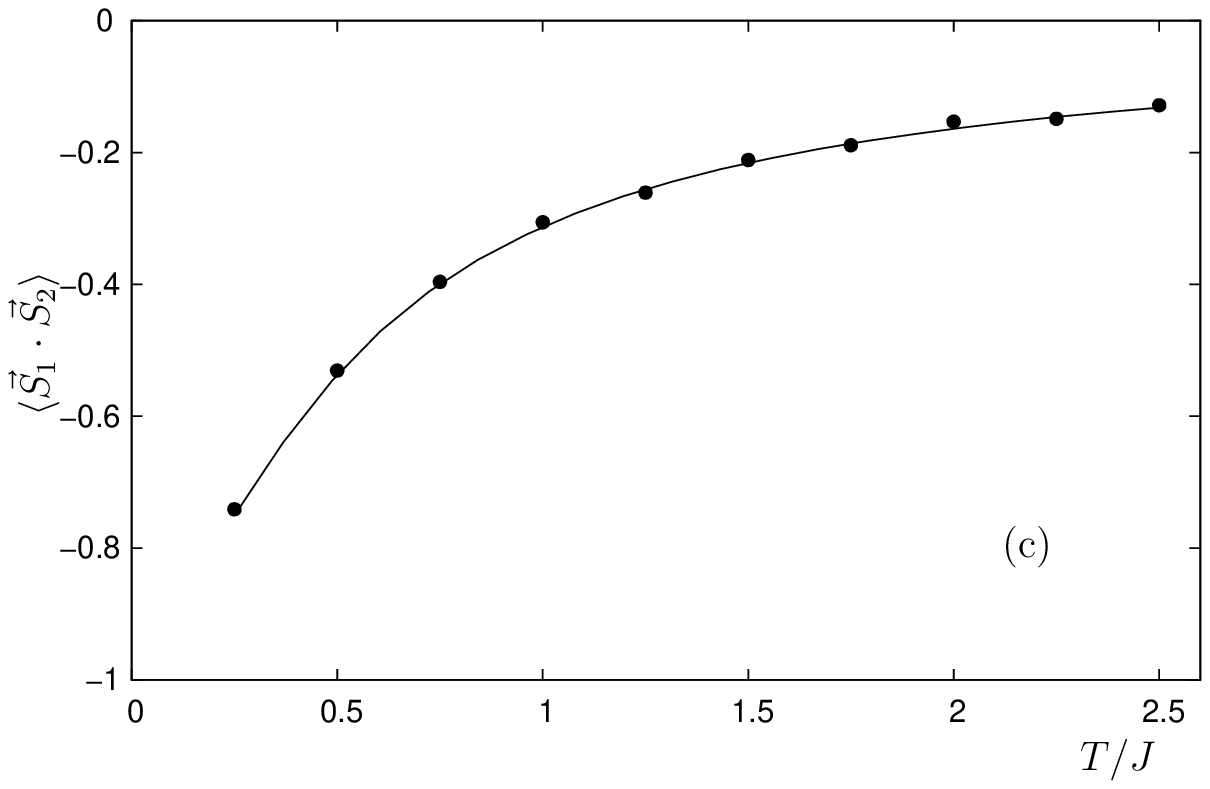,width=0.45\textwidth}
\epsfig{file=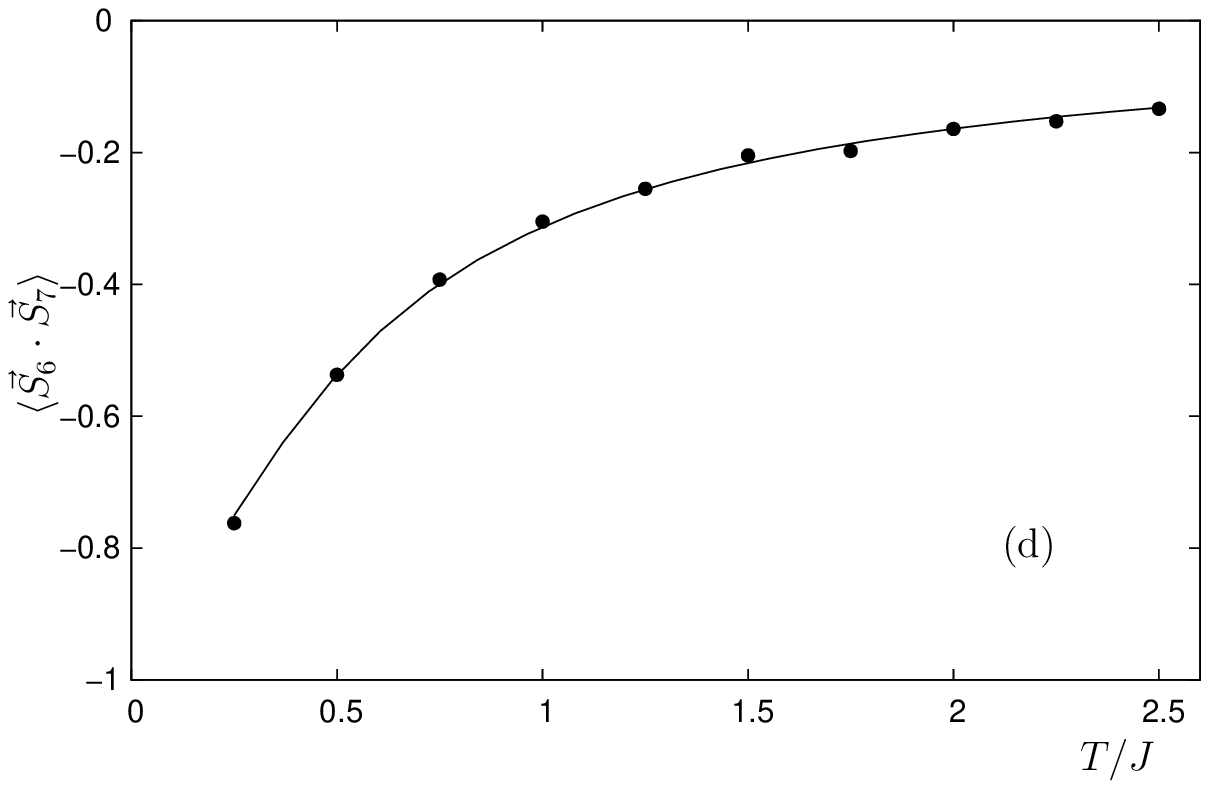,width=0.45\textwidth}
\end{center}
\caption{\label{fig:1} Correlation functions of local neighboring
spins in a finite 16-site chain, $\langle
\vec{S}_{g}\cdotp\vec{S}_{g+1}\rangle$, versus temperature in the
range $0<T<2.5 J$. The values of local spins are normalized to
$S=1$. The parameters of the Hamiltonian are $V=100J, t=50J$. The
panels corresponds to the different sites (a) $g=-1$, (b) $g=0$,
(c) $g=1$, and (d) $g=6$. The plots for sites $g=2, \dots, 6$ are
nearly the same. Also notice that the chain is symmetric with
respect to the impurity located at $g=-1/2$. In all panels, solid
lines correspond to the variational method described in the text,
and points to the MC simulations.}
\end{figure}

To analyze the evolution of the magnetic polaron state with the
growth of temperature, we calculate the correlation functions of
neighbouring spins in the chain, i.e. mean values $\langle
\vec{S}_{g}\cdotp\vec{S}_{g+1}\rangle=S^{2}\langle\cos\nu_g\rangle$.
For the temperature range $0<T<2.5 J$, we have performed the
calculation of correlations function using both methods described
above. Corresponding curves are shown in Fig.~\ref{fig:1} for
$V=100J, t=50J$. It can be seen that both methods give similar
results. Figure~\ref{fig:1} shows that for the four sites closest to the
impurity, we have ferromagnetic correlations, i.e.
$\langle\cos\nu_{g}\rangle >0$ for $g=-2,-1,0$. The remaining part
of the chain exhibits antiferromagnetic correlations. At zero
temperature, local spins inside the ferron are nearly parallel
each other, i.e. $\langle\cos\nu_g\rangle\approx 1$, whereas
neighboring spins in the remaining part of the chain are antiparallel,
i.e. $\langle\cos\nu_g\rangle\approx -1$.

Due to one-dimensionality of the model, the magnetic phase
transitions in the chain do not occur. Nevertheless, we can
introduce some characteristic temperatures describing our system.
The antiferromagnetic correlation functions,
$\langle\cos\nu_{g}\rangle$, between neighboring spins far from
the impurity exhibit similar behavior. We may define temperature
$T^{*}_{AF}$ corresponding to the N\'eel temperature of 3D
Heisenberg antiferromagnet in a conventional way, as the point of
the steepest change in this correlation function (that is the
point where its curvature has a maximum). This temperature does
not depend on $V$, and equals approximately to $T^{*}_{AF}=0.56J$,
which is close to the mean-field estimate of this parameter. From
Fig.~\ref{fig:1}, we see that it corresponds to the point where
$\langle\cos\nu_{6}\rangle|_{T=T^{*}_{AF}}\approx -1/2$.

The plots in Fig.~\ref{fig:1} demonstrate that the ferromagnetic
correlations are significant for the  first and second magnetic
neighbors to the impurity, even at high temperature. The AF
correlations in the rest part of the chain decay much faster with
temperatures. So, the ferron is a stable object that does not
disappear even at $T \gg T^{*}_{AF}$.

\begin{figure}
\begin{center}
\epsfig{file=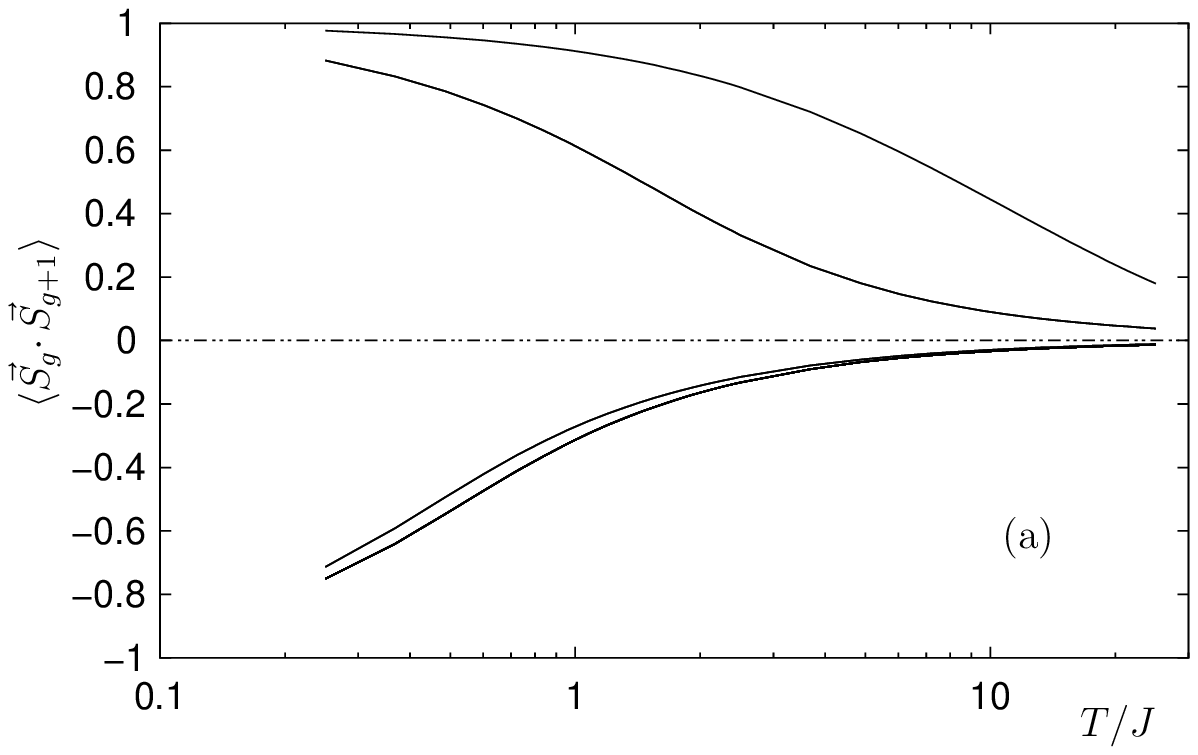,width=0.6\textwidth}\\
\epsfig{file=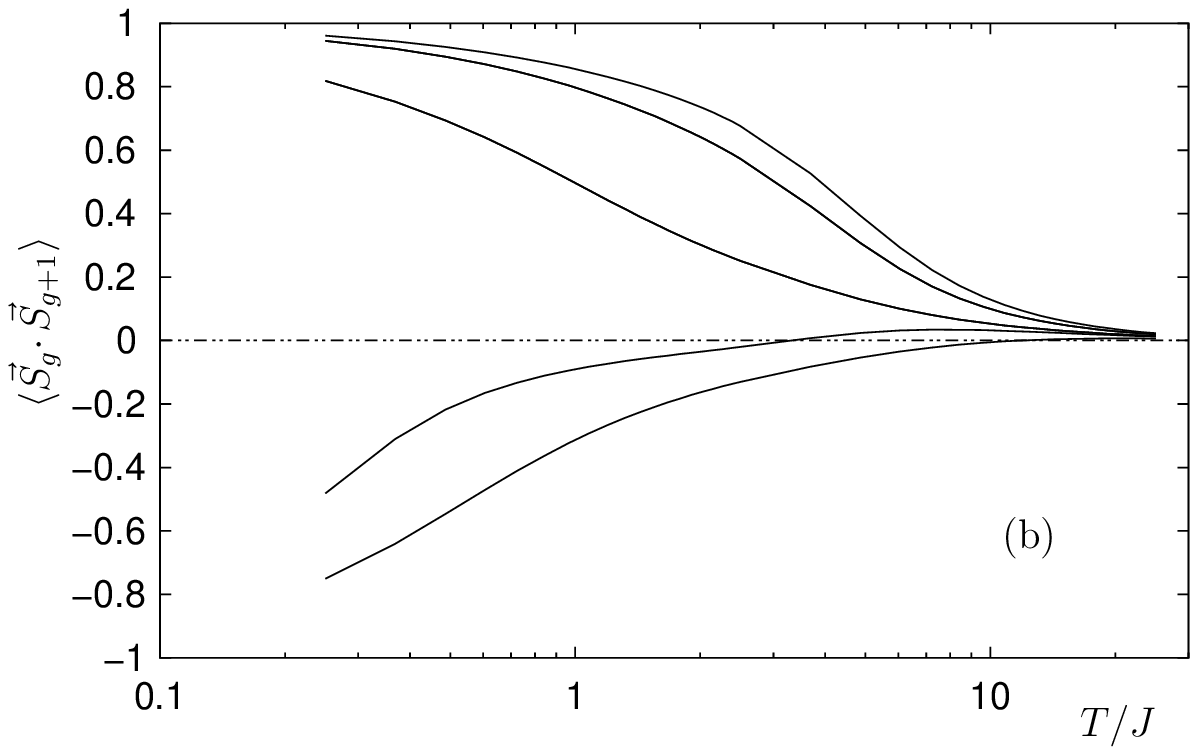,width=0.6\textwidth}
\end{center}
\caption{\label{fig:2} Correlation functions
$\langle\vec{S}_{g}\cdotp\vec{S}_{g+1}\rangle=\langle\cos\nu_g\rangle$
versus temperature at parameters $V=100J$, $t=50J$ (a), and
$V=5J$, $t=50J$ (b). In both cases, the upper curve corresponds to
the correlation between spins at sites near to the impurity
$\langle\cos\nu_{-1}\rangle$, the next one is the correlation
between next two spins $\langle\cos\nu_{0}\rangle$ and so on. The
lower curve is the correlation between last two spins in the chain
$\langle\cos\nu_6\rangle$. In panel (a), ferromagnetic
correlations exist for first two sites near to the impurity, and
magnetic polaron contains $4$ sites, whereas at smaller $V$ (panel
(b)), it contains $6$ sites.}
\end{figure}

Let study the temperature range in which the ferromagnetic correlations
start to decay steeply (see Fig.~\ref{fig:2}). We introduce, similarly to
$T^{*}_{AF}$, a set of temperatures characterizing the magnetic polaron:
$T^{*}_{1}, T^{*}_{2}, \dots, T^{*}_{n}$, $n$ being the last site showing
ferromagnetic correlations at zero temperature. Namely, $T^{*}_{1}$ is
the temperature corresponding to the maximum curvature of the plot shown
in Fig.~\ref{fig:2} for the correlation function for the sites nearest to
the impurity. From Fig.~\ref{fig:2}, we see that this temperature
corresponds to the value $\langle\cos\nu_{-1}\rangle|_{T=T^{*}_1}\approx+1/2$.
Similarly, $T^{*}_{2}$ corresponds to the value at which
$\langle\cos\nu_0\rangle|_{T=T^{*}_2}\approx +1/2$, and so on. In the case of
the sets of parameters corresponding to Fig.~\ref{fig:2}, $n=2$
for $V=100J$, and $n=3$ for $V=5J$. These temperatures are
$T^{*}_2=1.45J$, $T^{*}_1=8.34J$ for $V=100J$, $t=50J$,
and $T^{*}_3=J$, $T^{*}_2=3.08J$, $T^{*}_1=3.92J$ for $V=5J$, $t=50J$.
At temperature $T=T^{*}_2$ in the case of $V=100J$, and at $T=T^{*}_3$
for $V=5J$, the ferromagnetic correlations inside the ferron start
to decay, and at $T>T^{*}_1$, the ferron state completely disappears.

The calculation shows that the ferron state is more stable at
larger values of $V$. However, this dependence is not very much
pronounced: the change in the characteristic temperatures is by a
factor of $2$, when $V$ changes from $100J$ to $5J$. The
temperature $T^{*}_{1}$, which characterizes the complete decay of
the ferron, is in the range of $0.1-0.2 t$. The value of
$T^{*}_{1}$ can be obtained on the basis of following simple
estimates. If we note that at $T^{*}_{1}$ all correlation
functions except the first one go to zero, and that the wave
function of the electron does not change very much with
temperature, we can approximate the gain in energy of having a
ferron state as $\langle\sum\limits_{g}
t\psi_{g}\psi_{g+1}\cos\left(\nu_{g}/2\right)\rangle$. Using that
at $T=T^{*}_{1}$, $\langle\cos\nu_{-1}\rangle\approx 1/2$, and
assuming that the wave function is uniformly distributed over the
size of ferron, we have $|\psi_{g}|^{2}=1/4$ for the $4$-site
ferron,  $|\psi_{g}|^{2}=1/6$ for the $6$-site ferron. Then,
$T^{*}_{1}\sim \langle t|\psi_{0}|^{2}\cos\left(\nu_{-1}/2\right)
\rangle\approx 0.1\,t-0.2\,t$. Also note that the relationship
between $T^{*}_{1}$ and $t$ is similar to the relationship between
the Curie temperature $T_{\text{C}}$ and $t$ for the double
exchange model. This similarity could have the same physical
nature since in the latter model, the ferromagnetic ordering is
related to the same hoping integral multiplied by the density of
conduction electrons on a lattice site~\cite{izy01,dag03b,yun98}.

\begin{figure}
\begin{center}
\epsfig{file=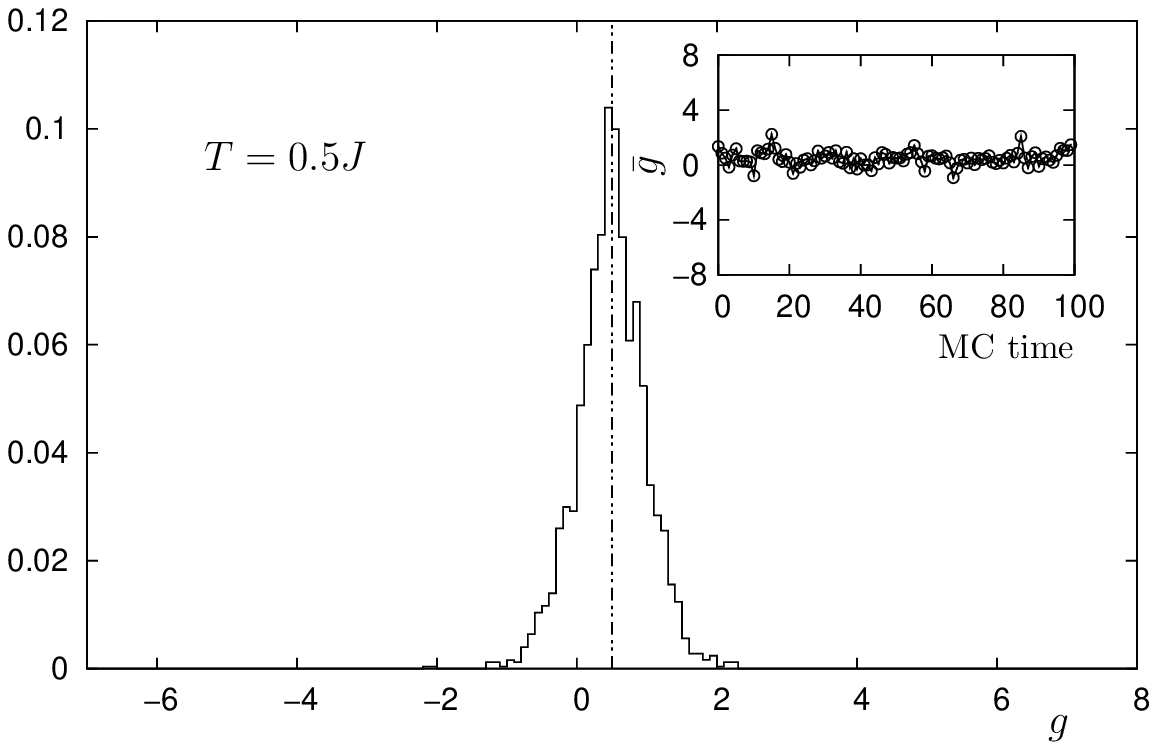,width=0.44\textwidth,height=0.19\textheight}
\epsfig{file=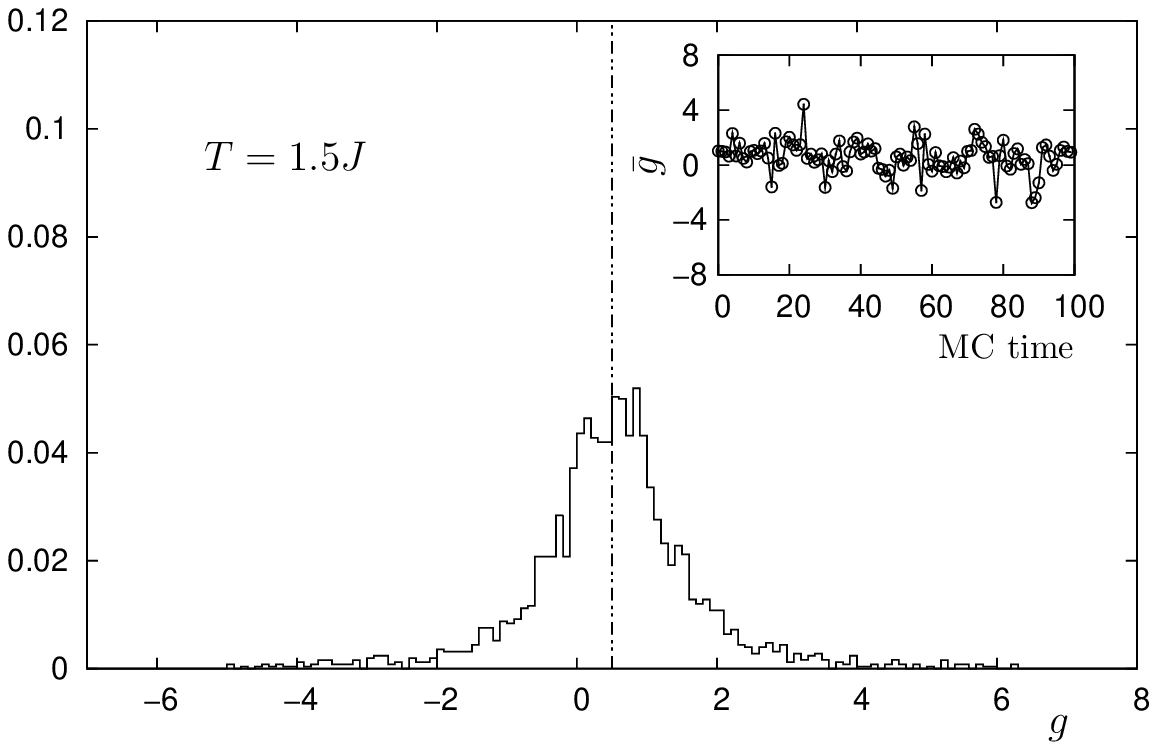,width=0.44\textwidth,height=0.19\textheight}\\
\epsfig{file=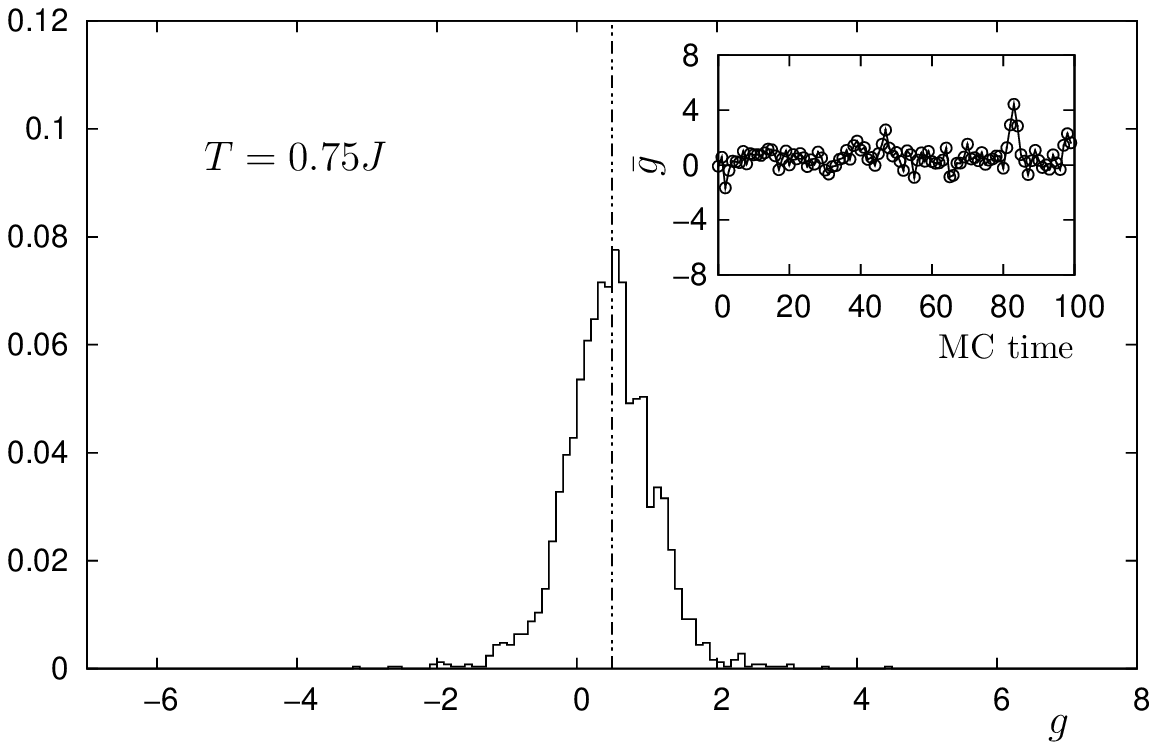,width=0.44\textwidth,height=0.19\textheight}
\epsfig{file=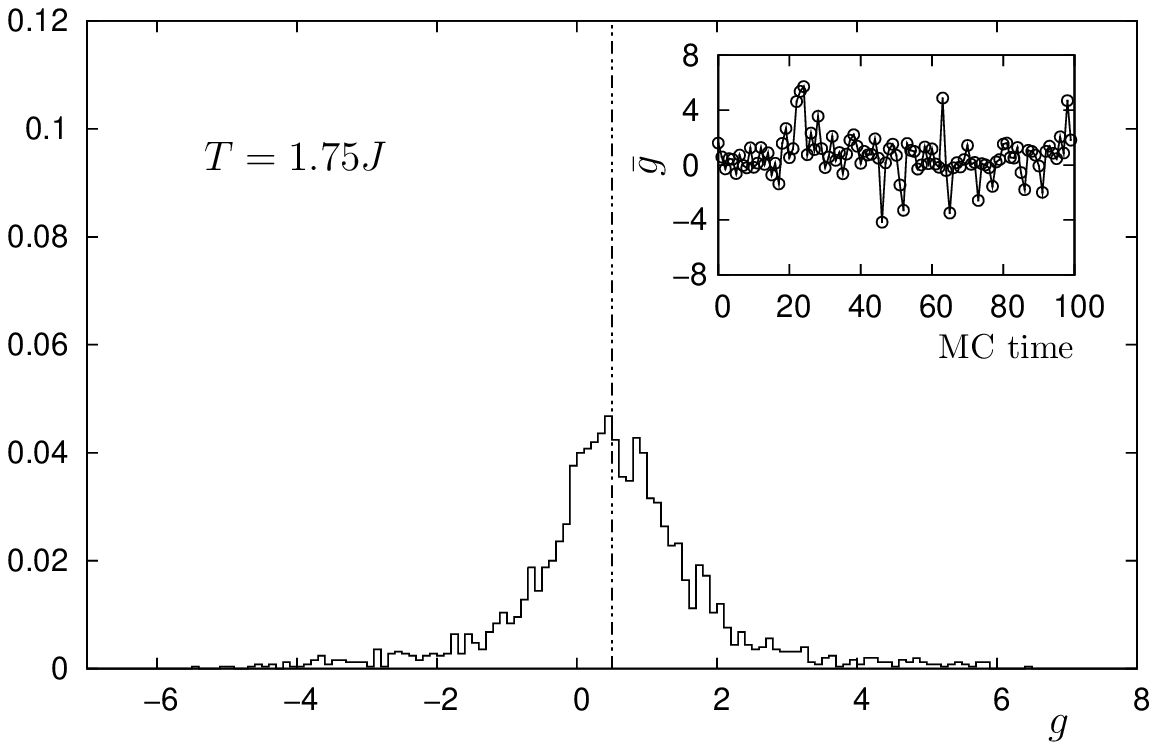,width=0.44\textwidth,height=0.19\textheight}\\
\epsfig{file=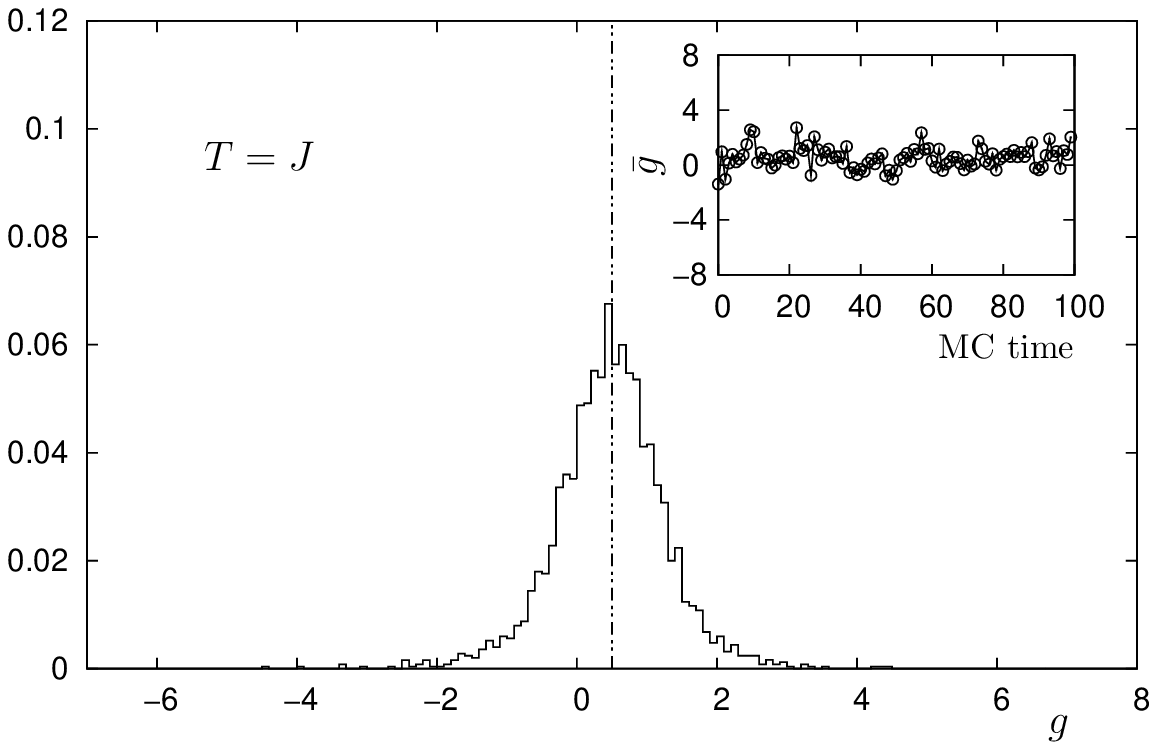,width=0.44\textwidth,height=0.19\textheight}
\epsfig{file=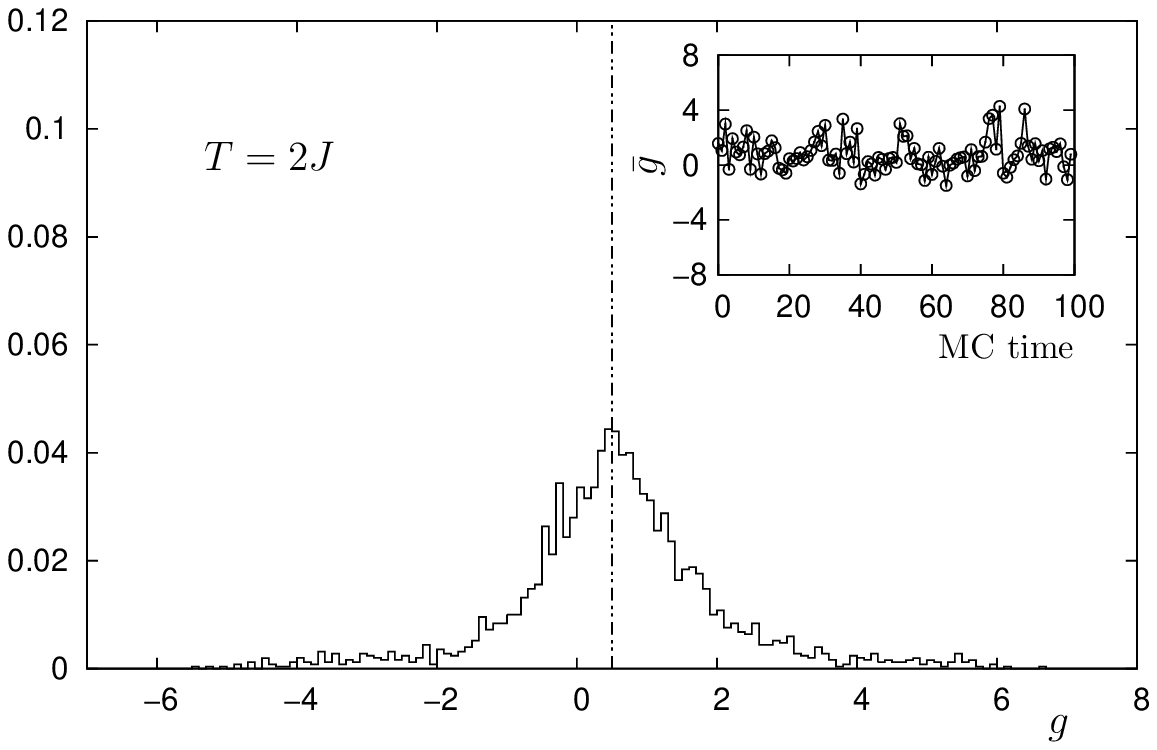,width=0.44\textwidth,height=0.19\textheight}\\
\epsfig{file=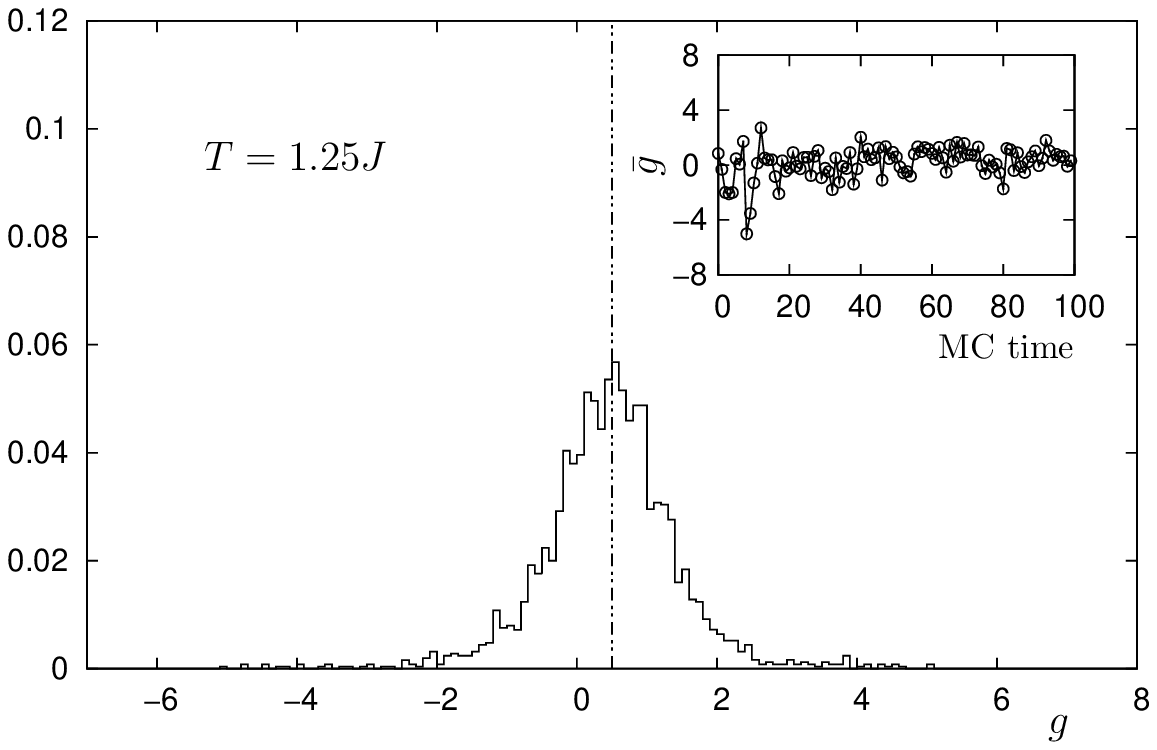,width=0.44\textwidth,height=0.19\textheight}
\epsfig{file=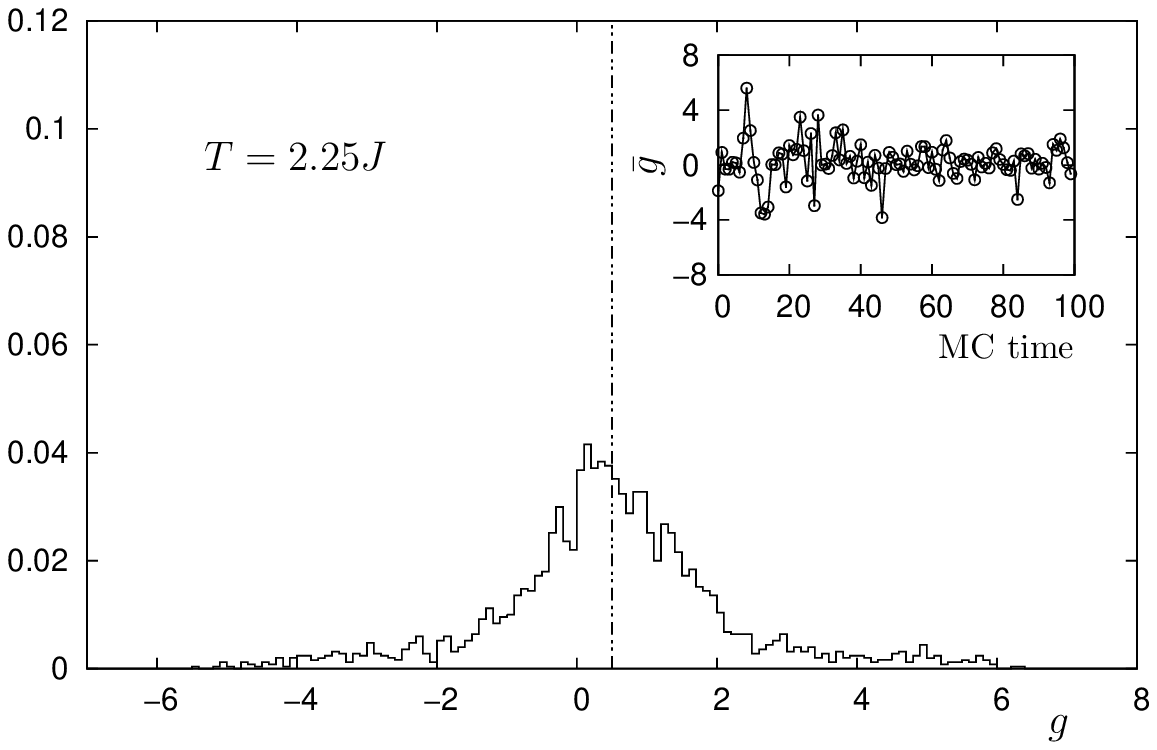,width=0.44\textwidth,height=0.19\textheight}
\end{center}
\caption{\label{fig:3}  Probability distribution of the
quantum-mechanical mean position of the conduction electron,
$\bar{g}$, for different values of temperature, and $V=5\,J$. The
probability distribution broadens with the growth of temperature.
The insets show the value of $\bar{g}$ as a function of the MC
time. At $T\geq1.5\,J$ there is a large scatter of $\bar{g}$,
which is the signature of the ferron depinning. Vertical
dot-dashed line indicates the impurity position.}
\end{figure}

Using our approximate variational method, we can analyze the case
of $V\rightarrow 0$. The plots for the correlation functions are
very similar to those in of Fig.~\ref{fig:2}b. In this limit, we
have $n=4$, so the ferron contains $8$ sites. The values of the
corresponding characteristic temperatures are $T^{*}_{1}=2.97J$,
$T^{*}_{2}=2.57J$, $T^{*}_{3}=1.31J$, and $T^{*}_{4}=0.14J$.
Therefore, in our approach, we can describe both the ferron bound
by impurity, and depinned from it. Some indication of this
possible depinning at small values of $V$ could come from the
presence of small ferromagnetic correlation corresponding to the
fourth pair at high temperatures (see Fig.~\ref{fig:2}b).

To study the possible depinning in a more detailed way, we
calculate the quantum-mechanical mean position
$\bar{g}=Z^{-1}_{\text{el}}\sum\limits_{n}\sum\limits_{g}\langle
\psi_g^{(n)}|\,g\,\psi_g^{(n)}\rangle\,
e^{-\beta E_{n}}$ of the conduction electron for each MC time, where
$Z_{\text{el}}=\sum\limits_{n}e^{-\beta E_{n}}$.
Dependence of mean position $\bar{g}$ on MC time is shown in
insets to Fig.~\ref{fig:3}. In this figure, we also show the
histograms illustrating probability distribution of $\bar{g}$ in
the temperature range $0<T<2.5J$, and for the set of parameters
$V=5J, t=50J$. In the insets we can see that at temperatures
$T\geq1.5\,J$, there appears a large scatter in the values of mean
position $\bar{g}$. Simultaneously, the probability distribution
broadens. This means that an electron becomes depinned and can
move far away from the impurity. Since the depinning temperature
is, at least, lower than the temperatures $T^{*}_{1}, T^{*}_{2}$
for the same values of $V$, and $t$, we can conclude that the
ferron retains its magnetic structure after depinning. At
sufficiently large values of $V$, the depinning of a ferron is not
observed in the temperature range under study.

\begin{figure}
\begin{center}
\epsfig{file=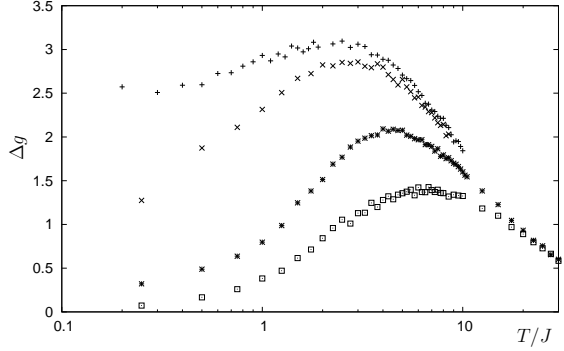,width=0.45\textwidth}
\end{center}
\caption{\label{fig5}  Standard deviation of $\bar{g}$, $\Delta
g$, as a function of $T$. Different curves correspond to different
impurity potentials: $V=0.1,\,1,\,5,$ and $100\,J$ (corresponding
the first to the highest, and the last to the lowest curve,
respectively).}
\end{figure}

In Fig.~\ref{fig5}, we plot the standard deviation of the histograms of
$\bar{g}$ as a function of temperature for several values of $V$. In
all cases, the standard deviation has a maximum for some
$T_{\text{max}}>J$.
At relatively low temperatures, $\Delta g$ grows due to the
increase in the thermal fluctuations of the ferron position. At
higher temperatures, the ferromagnetic correlations forming the
ferron gradually decay, and the electron undergoes a kind of
Anderson localization in disordered spin background (the situation
typical of the double exchange mechanism). Therefore, $\Delta g$
eventually decreases independently of the value of Coulomb
potential $V$.

We can also find the depinning line using the variational method.
Let us consider the ferron state bound to the impurity and the
state at $V\to0$. The latter case corresponds also to the free
ferron located far from the impurity. The energy difference
between bound and free ferron states then equals to
$E(0,T_{dp})-E(V,T)$. The depinning temperature $T_{dp}$ can be
estimated as:
\begin{equation}
E(0,T_{dp})-E(V,T_{dp})=T_{dp}\,,
\end{equation}
where the mean energy of the system $E(V,T)$ is calculated by
variational method using formula~\eqref{meanE}. Note that the
depinning temperature calculated in such a way turns out to be in
several times greater than its value estimated by Monte Carlo
simulations, because in latter case we consider the states
corresponding to the ferron located not far from the impurity.

\begin{figure}
\begin{center}
\epsfig{file=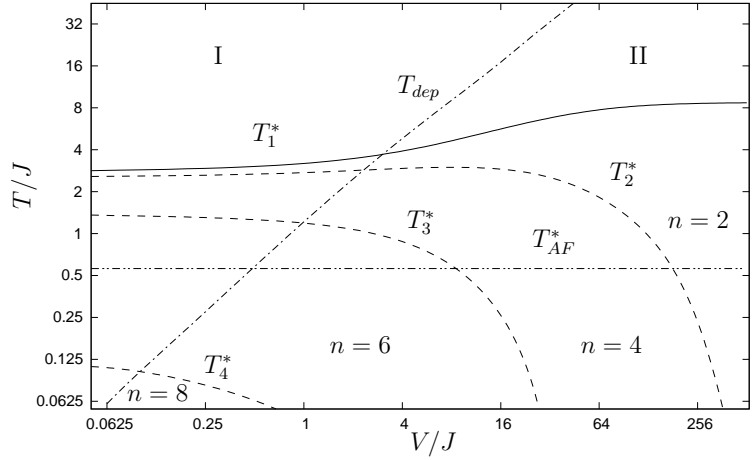,width=0.6\textwidth}
\end{center}
\caption{\label{fig:4}  Phase diagram in the $V-T$ plane for a
chain with a low density of bound ferrons. The value $t=50{J}$ is
fixed. Solid line corresponds to $T^{*}_{1}$, that is, to the
complete decay of the ferron. Dashed lines are the $T^{*}_{n/2}$
plots, depicting the decay of ferromagnetic correlations at the
end pair of spins in the ferron. Horizontal line corresponds to
the temperature $T^{*}_{AF}$. Dot-dashed line is the depinning
line obtained by estimations described in the text. Above this
line, ferrons are not bound by the impurities. Regions I and II above the
``melting'' line $T^{*}_{1}$ correspond to the free electron in the chain,
and to the electron bound by the impurity in the absence of the ferron
state.}
\end{figure}

In Fig.~\ref{fig:4}, we present the phase diagram of the chain in
$V-T$ plane. The temperatures $T^{*}_{1}, T^{*}_{2}, \dots,
T^{*}_{n}$ as a functions of the impurity potential are shown. We
also plot the temperature $T^{*}_{AF}$ and the depinning line
calculated by the variational method.

\section{Conclusions}

We show that the simple model of an antiferromagnetic chain with
impurities captures the essential features of the phase separation
in low-doped manganites. We find a set of temperatures
characterizing the thermal evolution of the magnetic structure of
such a chain. We demonstrate that the magnetic state of the chain
is rather inhomogeneous and it can be described by a set of
magnetic polarons. These magnetic polarons are rather stable
objects existing at temperature even much higher than the N\'eel
temperature of the undoped chain. With the growth of temperature,
ferromagnetic correlations inside the ferron start to disappear in
a step-wise way, up to some temperature $T^{*}_{1}\sim
0.1\,t-0.2\,t$, and then ferron state is no longer stable. The
dynamics of ferrons depends on the magnitude of the
electron-impurity coupling, $V$. At high values of $V$, ferron
remains tightly bound by impurity, while at lower $V$, the ferron
can be depinned from impurity at a characteristic temperature
lower than $T^{*}_{1}$, and therefore retains its magnetic
structure after depinning. These results can be summarized in the
phase diagram in the $V-T$ plane presented in Fig.~\ref{fig:4}. We
can see that in region below the rather flat curve $T^{*}_{1}$
versus $V$, there are the regions of stability of different kinds
of ferrons, from $n=2$ to $8$ sites. In each such region, we show
the dashed line corresponding to the melting of the last
ferromagnetically correlated pair in the ferron. So, between the
solid and each dashed line, we have the region where the ferron
gradually decays. We also show the depinning line calculated by
variational method as it was discussed at the end of the previous
section. We see that there is a large region between the depinning
line and $T^{*}_{1}$ where ferron is depinned. We have extended
this depinning line above the $T^{*}_{1}$ curve to show the
possible domains of existence of bare electron bound to the
impurity and freely moving along the chain.

\section*{Acknowledgments}

This work was supported by the Russian Foundation for Basic
Research (projects 02--02--16708 and NSh-1694.2003.2).
K.~I. Kugel and I. Gonz\'alez also acknowledge the support
from Xunta de Galicia and Universidade de Santiago de Compostela.

\end{document}